\documentclass[draftcls,12pt,onecolumn]{IEEEtran}
\usepackage{amsfonts}
\usepackage{amssymb}
\usepackage{mathrsfs}
\usepackage{verbatim}
\usepackage{subfigure}
\usepackage[centertags]{amsmath}
\usepackage{graphicx}  
\newtheorem{theorem}{\bf Theorem}
\newtheorem{construction}{ Construction}[section]

\newtheorem{definition}{Definition}

\newtheorem{example}{Example}[section]

\title{MMSE Optimal Algebraic Space-Time Codes}
\author{G. Susinder Rajan and B. Sundar Rajan
\thanks{This work was supported through grants to B.S.~Rajan; partly by the DRDO-IISc program on Advanced Research in Mathematical Engineering, and partly by the Council of Scientific \& Industrial Research (CSIR, India) Research Grant (22(0365)/04/EMR-II). Part of the material in this letter has been published in the Proceedings of Thirteenth National Conference on Communications (NCC 2007) held at IIT Kanpur, January 27-29, 2007. G. Susinder Rajan and B. Sundar Rajan are with the Department of Electrical Communication Engineering, Indian Institute of Science, Bangalore-560012, India. Email:\{susinder,bsrajan\}@ece.iisc.ernet.in.}}

\begin{document}
\maketitle
\begin{abstract}
Design of Space-Time Block Codes (STBCs) for Maximum Likelihood (ML) reception has been predominantly the main focus of researchers. However, the ML decoding complexity of STBCs becomes prohibitive large as the number of transmit and receive antennas increase. Hence it is natural to resort to a suboptimal reception technique like linear Minimum Mean Squared Error (MMSE) receiver. Barbarossa \emph{et al} and Liu \emph{et al} have independently derived necessary and sufficient conditions for a full rate linear STBC to be MMSE optimal, i.e achieve least Symbol Error Rate (SER). Motivated by this problem, certain existing high rate STBC constructions from crossed product algebras are identified to be MMSE optimal. Also, it is shown that a certain class of codes from cyclic division algebras which are special cases of crossed product algebras are MMSE optimal. Hence, these STBCs achieve least SER when MMSE reception is employed and are fully diverse when ML reception is employed.  
\end{abstract}
\begin{keywords}
Crossed product algebra, division algebra, space-time codes, MMSE receiver
\end{keywords}
\section{Introduction}

Space-Time coding is known to be an efficient coding technique to combat fading and/or exploit the increased capacity gains offered by Multiple Input Multiple Output (MIMO) systems. But the ML decoding complexity of STBCs becomes prohibitively large as the number of transmit and receive antennas increase. The sphere decoder helps to some extent in reducing the complexity but is still far away from practicality for large number of transmit antennas. In \cite{TJC,Jaf,KhR}, orthogonal designs, single and double symbol ML decodable STBCs have been proposed to solve this problem. But unfortunately, the rate of such codes decay with increase in the number of transmit antennas and they are information lossy for more than one receive antenna. This led to the study of suboptimal reception strategies such as linear MMSE (Minimum Mean Square Error) and linear ZF (Zero Forcing) receivers \cite{LZW1}-\cite{ZLW2}. It is then natural to address the question of how to design STBCs which are optimal for a linear MMSE receiver. This problem was addressed in \cite{LZW1}-\cite{BaF}.

\label{sec1}
\begin{definition}
\label{defn_mmse_optimal}
A $n \times n$ linear STBC $S$ in $k$ complex variables $x_1,\dots,x_k$ given by $S  = \sum_{i=1}^{k} x_{i}A_{i}$ is called a unitary trace-orthogonal STBC if the set of $n\times n$ matrices $A_i,i=1,\dots,k$ satisfy the following conditions
\begin{eqnarray}
\label{unitary}
A_iA_i^H&=&\frac{n}{k}I_n\\ 
\label{torthogonal}
Tr(A_i^HA_j)&=&0, \forall\ i\neq j
\end{eqnarray}
If $k=n^2$, it will be referred to as full rate transmission. 
\end{definition}
It was shown in \cite{LZW1}-\cite{BaF} that if full rate transmission is considered, unitary trace-orthogonality is a necessary and sufficient condition for a linear STBC to achieve minimum bit error rate when the variables $x_1,\dots,x_k$ take values from a QPSK (Quadrature Phase Shift Keying) constellation. Further, it was shown that full rate unitary trace orthogonal STBCs achieve MMSE when other two-dimensional constellations are used. Also, it was shown that at high SNR, the predominant metric that decides probability of symbol error is optimized only by unitary trace orthogonal STBCs. Henceforth, we thus refer to full rate unitary trace orthogonal STBCs as MMSE optimal STBCs. Few constructions of such codes are given in \cite{LZW2}-\cite{Bar}. However, these constructions were based on matrix manipulations and lacked an algebraic theory behind them.

The contributions of this paper are as follows.
\begin{itemize}
\item Provide sufficient conditions as to when STBCs obtained from left regular representation of crossed product algebras are MMSE optimal. Using these sufficient conditions, a new class of MMSE optimal STBCs is constructed for arbitrary number of transmit antennas. Since the code constructions are algebraic, the description of the code becomes elegant and it also simplifies the study of their properties. 
\item By restricting to a certain class of cyclic division algebras \cite{SRS1}, STBCs which are simultaneously MMSE optimal as well as fully diverse for ML reception are identified. Not all division algebra based codes \cite{SRS1}-\cite{EKPKL} are MMSE optimal. In particular, it is shown that the famous Golden code \cite{ORBV} is not MMSE optimal. Few of the existing code constructions \cite{LZW2,Jing_thesis,Bar} are also shown to be special cases of certain codes from cyclic algebras \cite{SRS2,SRS1}.
\end{itemize}

\subsection{Organization of the paper}
In Section \ref{sec2}, a description of our main algebraic tool, i.e., crossed product algebras is provided and an explicit construction of STBCs from crossed product algebras is given. In Section \ref{sec3}, we identify sufficient conditions as to when STBCs from crossed product algebras are MMSE optimal. Then, we focus on a proper subclass of crossed product algebras called cyclic algebras and it is shown that a certain class among them are MMSE optimal as well. Few illustrative examples of code constructions are provided and the decoding procedure for these codes is briefly discussed. Simulation results comprise Section \ref{sec4} and discussions on future work constitute Section \ref{sec5}.    

\section{STBCs from Crossed Product Algebras}
\label{sec2}
In this section, we briefly review the construction of STBCs from crossed product algebras as given in \cite{SRS2}. We refer the readers to \cite{SRS2} for a detailed explanation of crossed product algebras.

Let $F$ be a field. Then, an associative $F$-algebra $A$ is called a central simple algebra if the center of $A$ is $F$ and $A$ is a simple algebra, i.e., $A$ does not have nontrivial two-sided ideals. Simple examples of central simple algebras are division algebras and matrix algebras over fields. It is well known that the dimension $[A:F]$ of $A$ over its center is always a perfect square, say $n^2$ \cite{SRS2,Herstein}. The square root of $[A:F]$ is called the degree of $A$. Let $K$ be a strictly maximal subfield of $A$, i.e., $K\subset A$ and $K$ is not contained in any other subfield of $A$ and the centralizer of $K$ in $A$ is $K$ itself. It is well known that $[K:F]=n$, the degree of the algebra. In addition, let the extension $K/F$ be a Galois extension and let $G=\left\{\sigma_0=1,\sigma_1,\sigma_2,\ldots \sigma_{n-1}\right\}$ be the Galois group of $K/F$. Let $\phi$ be a map from $G \times G$ to $K$\textbackslash$\left\{0\right\}$ called the cocycle which satisfies the cocycle condition as shown below:
$$
\phi(\sigma,\tau\gamma)\phi(\tau,\gamma)=\phi(\sigma\tau,\gamma)\gamma(\phi(\sigma,\tau)),~\forall \sigma,\tau,\gamma \in G.
$$
Then, the algebra $A$ is called a Crossed Product Algebra if
\begin{equation*}
A=\bigoplus_{\sigma_i \in G}u_{\sigma_i}K
\end{equation*} 
where, equality and addition are component-wise and where $u_{\sigma}$ are symbols such that i) $\sigma(k)=u_{\sigma}^{-1}ku_{\sigma}$ and ii) $u_{\sigma}u_{\tau}=u_{\sigma\tau}\phi(\sigma,\tau)$ for all $k \in K,\sigma,\tau \in G$. It is clear that $A$ can be seen as a right $K$-space of dimension $n$ over $K$. Also multiplication between two elements of $A$, say $a=\sum_{i=0}^{n-1}u_{\sigma_i}k_{\sigma_i}$ and $a'=\sum_{j=0}^{n-1}u_{\sigma_j}k_{\sigma_j}^{'}$ is given by
\begin{equation*}
\left(\sum_{i=0}^{n-1}u_{\sigma_i}k_{\sigma_i}\right)\left(\sum_{j=0}^{n-1}u_{\sigma_j}k_{\sigma_j}^{'}\right)=
\sum_{l=0}^{n-1}u_{\sigma_l}k_{\sigma_l}^{''}
\end{equation*}
where, $k_{\sigma_l}^{''}=\sum_{\sigma_i\sigma_j=\sigma_l}\phi(\sigma_i,\sigma_j)\sigma_j(k_{\sigma_i})k_{\sigma_j}^{'}$
We will denote this crossed product algebra $A$ by $\left(K,G,\phi\right)$. The field $K$ can be seen as an $n$-dimensional $F$-vector space. Let $B= \left\{ t_0,t_1,\dots t_{n-1} \right\}$ be a basis of $K$ over $F$. Then, the left regular representation \cite{SRS2} of $A$ in $End_K(A)$\footnote{$End_K(A)$ denotes the set of all $K$ linear maps from $A$ to $A$.} is given by the map $L:A \mapsto End_K(A)$ which is defined as follows
\begin{equation*}
L\left(a\right)=\lambda_a\ \mathrm{where,}\ \lambda_a\left(u\right)=au,\forall u \in A.
\end{equation*}
The matrix representation $M_a$ of the linear transformation $\lambda_a$ with respect to the basis $\left\{u_{\sigma_i}:\sigma_i\in G\right\}$ is given by $\eqref{Ma}$ where, $f_{\sigma_j}^{(i)}\in F, \forall\ 0\leq i,j\leq n-1$, $\mu_{i,j}=\sigma_i\sigma_j^{-1}$, $\beta_i^{(j)}=\phi(\sigma_i\sigma_j^{-1},\sigma_j)$ and $\alpha$ is a scaling factor to normalize the average total power of a codeword to $n^2$. 
{\small
\begin{equation}
\label{Ma}
M_a=\frac{1}{\sqrt{\alpha}} \left[ \begin{array}{ccccc}
\sum_{i=0}^{n-1}f_{\sigma_{0}}^{(i)}t_i & \beta_0^{(1)}\sum_{i=0}^{n-1}f_{\mu_{0,1}}^{(i)}\sigma_1(t_i) & \beta_0^{(2)}\sum_{i=0}^{n-1}f_{\mu_{0,2}}^{(i)}\sigma_2(t_i) & \cdots & \beta_0^{(n-1)}\sum_{i=0}^{n-1}f_{\mu_{0,n-1}}^{(i)}\sigma_{n-1}(t_i)\\

\sum_{i=0}^{n-1}f_{\sigma_{1}}^{(i)}t_i & \beta_1^{(1)}\sum_{i=0}^{n-1}f_{\mu_{1,1}}^{(i)}\sigma_1(t_i) & \beta_1^{(2)}\sum_{i=0}^{n-1}f_{\mu_{1,2}}^{(i)}\sigma_2(t_i) & \cdots & \beta_1^{(n-1)}\sum_{i=0}^{n-1}f_{\mu_{1,n-1}}^{(i)}\sigma_{n-1}(t_i)\\

\vdots & \vdots & \vdots & \ddots & \vdots\\

\sum_{i=0}^{n-1}f_{\sigma_{n-1}}^{(i)}t_i & \beta_{n-1}^{(1)}\sum_{i=0}^{n-1}f_{\mu_{0,1}}^{(i)}\sigma_1(t_i) & \beta_{n-1}^{(2)}\sum_{i=0}^{n-1}f_{\mu_{0,2}}^{(i)}\sigma_2(t_i) & \cdots & \beta_{n-1}^{(n-1)}\sum_{i=0}^{n-1}f_{\mu_{0,n-1}}^{(i)}\sigma_{n-1}(t_i)\\

\end{array} \right]
\end{equation}
}
Thus we have obtained a full rate linear STBC $M_a$ in variables $f_{\sigma_j}^{(i)}, 0\leq i,j\leq n-1$ from the crossed product algebra $A$. $M_a$ can expressed in a linear dispersion form \mbox{$M_a=\sum_{j=0}^{n-1}\sum_{i=0}^{n-1}f_{\sigma_j}^{(i)}W_{i,j}$} where, the matrices $W_{i,j}$ are called the 'weight matrices' of $M_a$. Then, we have
\begin{equation}
\label{Wij}
W_{i,j}=\frac{1}{\sqrt{\alpha}}P_jQ_i,\ \mathrm{where}\ Q_i=\left[\begin{array}{cccc}
t_i & 0 & \cdots & 0\\
0 & \sigma_1(t_i) & \ddots & \vdots\\
\vdots & \ddots & \ddots & 0\\
0 & \cdots & 0 & \sigma_{n-1}(t_i) 
\end{array}
\right]
\end{equation}
and the matrix $P_j$ can be described as follows. Let us index the rows and columns of $P_j$ with the elements of $G$. Then the $(\sigma_k,\sigma_l)$-th entry of $P_j$ is equal to $\phi(\sigma_j,\sigma_l)$ if $\sigma_j\sigma_l=\sigma_k$ and $0$ otherwise. 

The matrices $P_j$ and $Q_i$ are nothing but the images of $u_{\sigma_j}$ and $t_i$ respectively under the map $L$. Note that the $P_j$ matrices are known as permutation matrices and are commonly used for group representation.
\section{MMSE Optimal STBCs}
\label{sec3}
In this Section, we identify sufficient conditions as to when STBCs from crossed product algebras are MMSE optimal. Then, we focus on a proper subclass of crossed product algebras called cyclic algebras and obtain a class of STBCs meeting the required conditions for MMSE optimality. Finally, the decoding procedure for the codes in this paper is discussed and its simplicity as compared to ML decoding is highlighted.  

\begin{theorem}
\label{thm_mmse}
The STBC $M_a$ constructed as shown in \eqref{Ma} using the crossed product algebra $A=\left(K,G,\phi\right)$ is MMSE optimal if
\begin{eqnarray}
\label{cond1}
\vert\sigma_j(t_i)\vert=\vert t_i\vert&=&\vert\phi(\sigma_i,\sigma_j)\vert=1,\forall\ 0\leq i,j \leq n-1\\  
\label{cond2}
\mathrm{and}\ \sum_{i=0}^{n-1}\sigma_j(t_i)(\sigma_{j'}(t_i))^{*}&=&0,\ \mathrm{if}\ j\neq j'. 
\end{eqnarray}
\end{theorem} 
\begin{proof}
We need to show that the weight matrices of $M_a$ satisfy \eqref{unitary} and \eqref{torthogonal}. Equation \eqref{cond1} implies that the matrices $P_j$ and $Q_i$ are scaled unitary matrices. The scaling factor $\alpha$ here equals $n$. Therefore $W_{i,j}W_{i,j}^H=\frac{I_n}{n}$ which implies \eqref{unitary} is satisfied.

It can be shown \cite{ZLW1} that the condition in \eqref{torthogonal} is equivalent to the condition that the matrix $\Phi$ as shown in \eqref{phi} satisfies $\Phi\Phi^H=nI_n^2$. 
\begin{equation}
\label{phi}
\Phi=\left[\begin{array}{ccccccc}vec(W_{0,0}) & vec(W_{1,0}) & \dots & vec(W_{n-1,0}) & vec(W_{0,n-1}) & \dots &  vec(W_{n-1,n-1})\end{array}\right]
\end{equation}
The $(k,l)$th element of $\Phi\Phi^H$ is given by $\sum_{a=0}^{n-1}\phi(\sigma_i\sigma_j^{-1},\sigma_j)\sigma_j(t_a)\left(\phi\left(\sigma_{i'}\sigma_{j'}^{-1},\sigma_{j'}\right)\sigma_{j'}(t_a)\right)^*$, which simplifies to $\phi(\sigma_i\sigma_j^{-1},\sigma_j)\phi\left(\sigma_{i'}\sigma_{j'}^{-1},\sigma_{j'}\right)\sum_{a=0}^{n-1}\sigma_{j}(t_a)(\sigma_{j'}(t_a))^*$ which is equal to zero from the statement of the theorem. If $k=l$, then we have $(\Phi\Phi^H)_{k,k}=\sum_{a=0}^{n-1}\vert\sigma_j(t_a)\vert^2=n$. Thus, $\Phi\Phi^H=nI_n^2$ which in turn implies \eqref{torthogonal} is satisfied.
\end{proof}
Theorem \ref{thm_mmse} gives conditions on the basis of a Galois extension and on the cocycle which result in MMSE optimal STBCs.

\subsection{STBCs from Cyclic Algebras}
In this subsection, using Theorem \ref{thm_mmse}, we identify an existing STBC construction \cite{SRS2,SRS1} based on cyclic algebras to be MMSE optimal.

An $F$-central simple algebra is called a cyclic algebra, if $A$ has a strictly maximal subfield $K$ which is a cyclic extension of the center $F$. Clearly, a cyclic algebra is a crossed product algebra. Let $\sigma$ be a generator of the Galois group $G$. If $u_{\sigma^i},i=0,1,\dots,n-1$ is a basis for the algebra $A$ over $K$, then we have 
\begin{equation*}
\begin{array}{rcl}
u_{\sigma^i}&=&u_{\sigma}^i\\
\mathrm{and}\ \phi(\sigma^i,\sigma^j)&=&\left\{\begin{array}{l}1,\quad \mathrm{if}\ i+j<n\\ \delta,\quad \mathrm{if}\ i+j\geq n\end{array}\right.
\end{array}
\end{equation*}
where, $u_{\sigma}^n=\delta$. Since the cocycle can now be described by just one element $\delta$ and similarly $G$ can be described by $\sigma$, we denote the crossed product algebra $(K,G,\phi)$ with $(K,\sigma,\delta)$. Thus, with $z=u_{\sigma}$, we have $A=(K,\sigma,\delta)=\bigoplus_{i=0}^{n-1}z^iK$ where, $z^n=\delta$ and $kz=z\sigma(k),\forall k\in K$. Note that if the smallest positive integer $t$ such that $\delta^t$ is the norm of some element in $K\backslash\left\{0\right\}$ is $n$, then the cyclic algebra $A=(K,\sigma,\delta)$ is a cyclic division algebra \cite{SRS1}.

\begin{construction}
\label{cons_CDA_trans}
Let $K/F$ be a cyclic extension of degree $n$ with $K=F(t_{n}=t^{1/n})$, $t,\omega_n\in F$, $\vert t\vert=1$. Here $\omega_n$ denotes the $n$th root of unity and $\sigma:t_n\mapsto\omega_nt_n$ is the generator of the Galois group. Let $\delta$ be a transcendental element over $K$. From Theorem \ref{thm_mmse}, the STBC arising from the cyclic division algebra $(K(\delta)/F(\delta),\sigma,\delta)$ is MMSE optimal since it satisfies the following identities
\begin{equation}
\begin{array}{c}
\vert t\vert=\vert \delta\vert=\vert\sigma^i(t_n)\vert=1,\ i=0,1,\dots,n-1\\
\mathrm{and}\ \sum_{i=0}^{n-1}(t_n)^i(\sigma^k(t_n^i))^*=0,\ \mathrm{if}\ k\neq 0.
\end{array}
\end{equation}
The MMSE optimal STBC $M_a$ is given by $M_a=\sum_{j=0}^{n-1}\sum_{i=0}^{n-1}f_{j}^{(i)}W_{i,j},~f_{j}^{(i)}\in F$
where, the weight matrix $W_{i,j}=t_n^iP^jQ^i$. The matrices $P$ and $Q$ are as shown below:
\begin{equation}
P=\left( \begin{array}{ccccc}
0 & \dots & \dots & 0 & \delta\\
1 & 0 & \dots & 0 & 0\\
0& 1 & \ddots & \vdots & \vdots\\
\vdots & \ddots & \ddots & 0 & \vdots\\
0 & \dots & 0 & 1 & 0
\end{array} \right),\ Q=\left( \begin{array}{ccccc}
1 & 0 & \dots & 0 & 0\\
0 & \omega_n & \ddots & 0 & \vdots\\
\vdots & \ddots & \omega_n^2 & \ddots & \vdots\\
\vdots & 0 & \ddots & \ddots & 0\\
0 & \dots & \ldots & 0 & \omega_n^{n-1}
\end{array} \right).
\end{equation}
\end{construction}

We would like to emphasize here that the codes in \cite{LZW2,Jing_thesis,Bar} can be obtained as a special case of the above construction by simply choosing $\delta=1$. If $\delta=1$ then the algebra $A$ will be a cyclic algebra but is not guaranteed to be a division algebra. Also, we would like to point out that there are cyclic division algebra based STBC constructions in the literature \cite{ORBV,KiR,EKPKL} which opt to carefully choose the element $\delta$ to be from $F^*$ (rather than transcendental as in Construction \ref{cons_CDA_trans}) for other benefits such as achieving the diversity-multiplexing gain tradeoff. Some of those codes are now known as perfect STBCs \cite{ORBV}. It is important to note that not all cyclic division algebra based codes satisfy \eqref{cond1} and \eqref{cond2}. In fact there exist perfect STBCs which are not MMSE optimal. A concrete example of such a code is the best known $2$ transmit antenna STBC for ML reception, i.e., the famous Golden code. This is illustrated in the following example. 
\begin{example}
The codewords of the Golden code are given by $\frac{1}{\sqrt{5}}\left[\begin{array}{cc}
\alpha(a+b\theta) & \alpha(c+d\theta)\\
i\bar{\alpha}(c+d\bar{\theta}) & \bar{\alpha}(a+b\bar{\theta})
\end{array}\right]$ where, $a,b,c,d\in\mathbb{Z}[j]$, $\theta=\frac{1+\sqrt{5}}{2}$, $\bar{\theta}=\frac{1-\sqrt{5}}{2}$, $\alpha=1+j(1-\theta)$, $\bar{\alpha}=1+j(1-\bar{\theta})$. The weight matrices of the Golden code are given as follows:
$$
\frac{1}{\sqrt{5}}\left[\begin{array}{cc}\alpha & 0\\0 & \bar{\alpha}\end{array}\right], \frac{1}{\sqrt{5}}\left[\begin{array}{cc}\alpha\theta & 0\\0 & \bar{\alpha}\bar{\theta} \end{array}\right], \frac{1}{\sqrt{5}}\left[\begin{array}{cc}0 & \alpha\\ j\bar{\alpha} & 0\end{array}\right], \frac{1}{\sqrt{5}}\left[\begin{array}{cc}0 & \alpha\theta\\ j\bar{\alpha}\bar{\theta} & 0\end{array}\right].
$$
Clearly, the weight matrices of the Golden code are not scaled unitary which is a necessary condition for MMSE optimality (see \eqref{unitary} of Definition \ref{defn_mmse_optimal}). This is because the crossed product algebra associated with the Golden code fails to satisfy \eqref{cond1}. Hence the Golden code is not MMSE optimal.
\end{example}

\begin{example}

\label{eg_sim}
This example illustrates our construction procedure for $n=2$. Let $F=\mathbb{Q}(j,t)$, where $t$ is transcendental over $\mathbb{Q}(j)$. Then $K=F(t_2=\sqrt{t})$ is a cyclic extension of $F$ of degree $2$. The generator of the Galois group is given by $\sigma:t_2\mapsto-t_2$. Let $\delta$ be any transcendental element over $K$. Then $(K(\delta)/F(\delta),\sigma,\delta)$ is a cyclic division algebra. For example, we can choose $t=e^{j}$ and $\delta=e^{j\sqrt{5}}$. Then, we have $M_a=\frac{1}{\sqrt{2}}\left[\begin{array}{lr}f_0^{(0)}+f_0^{(1)}t_2 & \delta(f_1^{(0)}-f_1^{(1)}t_2)\\f_1^{(0)}+f_1^{(1)}t_2 & f_0^{(0)}-f_0^{(1)}t_2  \end{array}\right]$.
\end{example}
\begin{example}
This is an example of a MMSE optimal code which is not obtainable from a cyclic division algebra. Let $n=4$ and $F=\mathbb{Q}(j,x,y)$ where $x$ and $y$ are two transcendental numbers independent over $\mathbb{Q}(j)$. We choose these transcendental numbers to lie on the unit circle. Then $K=F(\sqrt{x},\sqrt{y})$ is a Galois extension of $F$ with the Galois group $G=\langle\sigma_x,\sigma_y\rangle$, where $\sigma_x:\sqrt{x}\mapsto-\sqrt{x}$ and $\sigma_y:\sqrt{y}\mapsto-\sqrt{y}$. The cocyle $\phi$ is defined as follows:
$$
\begin{array}{c}
\phi(\sigma_x,\sigma_x)=\phi(\sigma_x\sigma_y,\sigma_x)=\delta_1,~\phi(\sigma_y,\sigma_y)=\phi(\sigma_x\sigma_y,\sigma_y)=\delta_2,\\
\phi(\sigma_x,\sigma_y)=1~\mathrm{and}~\phi(\sigma_x\sigma_y,\sigma_x\sigma_y)=\delta_1\delta_2.
\end{array}
$$
Then, the algebra $(K(\delta_1,\delta_2),G,\phi)=K(\delta_1,\delta_2)\oplus u_{\sigma_x}K(\delta_2,\delta_2)\oplus\ u_{\sigma_y}K(\delta_1,\delta_2)\oplus u_{\sigma_x}u_{\sigma_y}K(\delta_1,\delta_2)$ is a crossed product algebra where, $\delta_1,\delta_2$ are independent transcendental numbers over $K$. Also, we choose to pick $\delta_1$ and $\delta_2$ to lie on the unit circle. The matrix representation of this crossed product algebra will give rise to an MMSE optimal STBC and has codewords of the form $\frac{1}{\sqrt{\alpha}}\left[\begin{array}{lccr}k_{0,0} & \delta_2\sigma_y(k_{0,1}) & \delta_1\sigma_x(k_{1,0}) & \delta_1\delta_2\sigma_x\sigma_y(k_{1,1})\\
k_{0,1} & \sigma_y(k_{0,0}) & \delta_1\sigma_x(k_{1,1}) & \delta_1\sigma_x\sigma_y(k_{1,0})\\
k_{1,0} & \delta_2\sigma_y(k_{1,1}) & \sigma_x(k_{0,0}) & \delta_2\sigma_x\sigma_y(k_{0,1})\\
k_{1,1} & \sigma_y(k_{1,0}) & \sigma_x(k_{0,1}) & \sigma_x\sigma_y(k_{0,0})
\end{array}\right]$ where each $k_{i,j},0\leq i,j\leq 1$ is given by $k_{i,j}=f_{i,j}^{(0)}+f_{i,j}^{(1)}\sqrt{x}+f_{i,j}^{(2)}\sqrt{y}+f_{i,j}^{(3)}\sqrt{xy}$ and $f_{i,j}^{(l)}\in\mathbb{Q}(j)\subset F$.
\end{example}

\subsection{Decoding procedure}
In this subsection, the decoding procedure for the codes in this paper is briefly explained and its receiver simplicity compared to ML reception is highlighted.
 
Let the encoded matrix $X=\sum_{j=0}^{n-1}\sum_{i=0}^{n-1}f_{j}^{(i)}W_{i,j}$. Let the number of receive antennas be $m$. We assume that $m\geq n$ in the sequel otherwise there will be an error floor \cite{FaB} when linear MMSE reception is employed. The received matrix $Y$ can be expressed as $Y=HX+N$,~where $H$ is the channel matrix of size $m\times n$ and $N$ is the $m\times n$ matrix representing the additive noise at the receiver whose entries are i.i.d. $\mathcal{CN}(0,1)$. Then, the linear MMSE receiver can be implemented in its simplest form as a \textit{symbol-by-symbol decoder} \cite{FaB}, as described below:
\begin{equation}
\label{eqn_mmse}
\hat{f}_{j}^{(i)}=tr(W_{i,j}^HJY)
\end{equation}
with $J=(H^HH+\frac{1}{\rho}I_n)^{-1}H^H$ where, $\rho$ is the Signal to Noise ratio (SNR) or equivalently in this case it is the average energy of the complex constellation used. Computation of $\hat{f}_{j}^{(i)}$ is then followed by hard decision, i.e., it is decoded to the nearest point (in the sense of Euclidean distance) in the constellation. Note that the decoding complexity is \textit{linear} in the size of the signal set which is far less compared to the complexity of sphere decoding.
\section{Simulation Results}
\label{sec4}
\begin{figure}[p]
\centering
\includegraphics{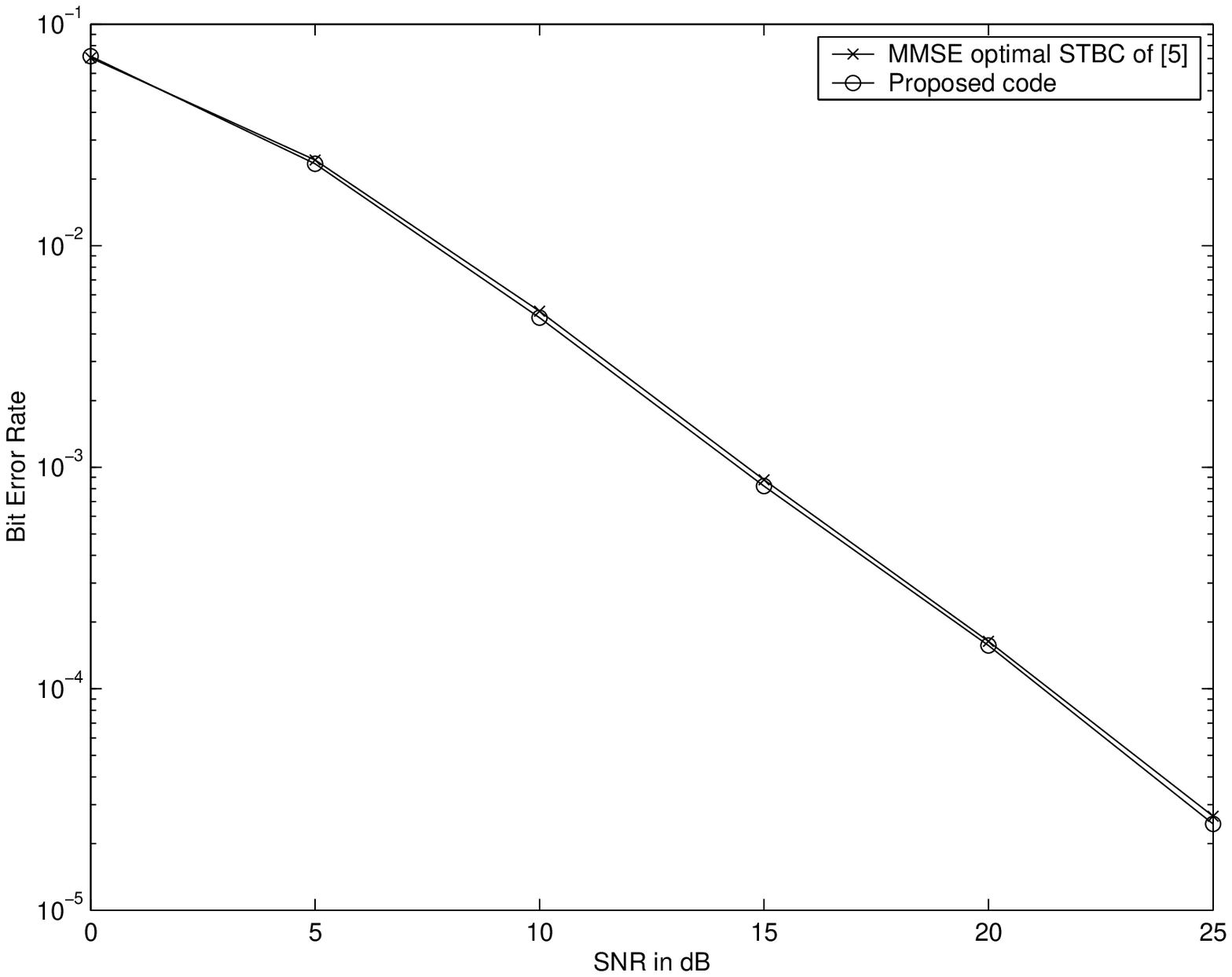}
\caption{Error performance comparison of the proposed MMSE optimal STBC with that of \cite{LZW2} in a $2\times 4$ MIMO system with a linear MMSE receiver}
\label{fig_mmse}
\end{figure}

In this section, we compare the error performance of the proposed codes with that of the previously known MMSE optimal STBC in \cite{LZW2} under linear MMSE reception. In \cite{FaB}, it has been shown that a diversity order of $m-n+1$ is achieved by MMSE optimal STBCs when a linear MMSE or linear Zero forcing receiver is employed. On the other hand, it is well known that under ML decoding a diversity order of $mn$ is possible if the STBC is fully diverse. The codes constructed in this letter have this property as well. Fig. \ref{fig_mmse} shows the bit error rate performance of the MMSE optimal STBC given in Example \ref{eg_sim} with QPSK constellation and the previously known MMSE optimal STBC in \cite{LZW2} under linear MMSE decoding with the number of receive antennas being equal to $4$. For linear MMSE decoding, the symbol-by-symbol decoder in \eqref{eqn_mmse} was utilized. Observe from Fig. \ref{fig_mmse} that the performance of the proposed code is almost same as that of the previously known MMSE optimal STBC in \cite{LZW2}. It is important to note that the error probability under linear MMSE reception as shown in Fig. \ref{fig_mmse} is the optimal \cite{LZW1}-\cite{BaF} among all STBCs with full rate transmission.

\section{Discussion}
\label{sec5}
The algebraic framework of crossed product algebras is quite general in nature. For instance, MMSE optimal STBCs can also be constructed from tensor products of division algebras and Brauer division algebras. We refer the readers to \cite{SRS2} for more details on these constructions. It will also be interesting to study the design of optimal STBCs for linear ZF receivers. Some initial work in this direction has been reported in \cite{ZLW2}.
\section*{Acknowledgements}
The authors thank Kiran and Shashidhar of Beceem Communications for useful discussions on the subject of this matter. The authors thank Jing Liu and Prof. Sergio Barbarossa for providing us with preprints of their recent works \cite{LZW2,Jing_thesis,ZLW2,FaB}. The authors are grateful to the anonymous reviewers for providing constructive comments and useful remarks on perfect space-time codes which helped in improving the presentation of this letter. 
\bibliographystyle{IEEEtran}

\end{document}